# Experimental optimization of lensless digital holographic microscopy with rotating-diffuser-based coherent noise reduction

PIOTR ARCAB[1,4], BARTOSZ MIRECKI[1], MARZENA STEFANIUK[2], MONIKA PAWŁOWSKA[2,3] AND MACIEJ TRUSIAK[1,*]

[1]*Warsaw University of Technology, Institute of Micromechanics and Photonics, 8 Sw. A. Boboli St., 02-525 Warsaw, Poland*
[2]*Laboratory of Neurobiology, BRAINCITY, Nencki Institute of Experimental Biology of Polish Academy of Sciences, Poland*
[3]*Institute of Experimental Physics, Faculty of Physics, University of Warsaw, Poland*
[4]*piotr.arcab.stud@pw.edu.pl*
*\*maciej.trusiak@pw.edu.pl*

**Abstract:** Laser-based lensless digital holographic microscopy (LDHM) is often spoiled by considerable coherent noise factor. We propose a novel LDHM method with significantly limited coherent artifacts, e.g., speckle noise and parasitic interference fringes. It is achieved by incorporating a rotating diffuser, which introduces partial spatial coherence and preserves high temporal coherence of laser light, crucial for credible in-line hologram reconstruction. We present the first implementation of the classical rotating diffuser concept in LDHM, significantly increasing the signal-to-noise ratio while preserving the straightforwardness and compactness of the LDHM imaging device. Prior to the introduction of the rotating diffusor, we performed LDHM experimental hardware optimization employing 4 light sources, 4 cameras and 3 different optical magnifications (camera-sample distances). It was guided by the quantitative assessment of numerical amplitude/phase reconstruction of test targets, conducted upon standard deviation calculation (noise factor quantification) and resolution evaluation (information throughput quantification). Optimized rotating diffuser LDHM (RD-LDHM) method was successfully corroborated in technical test target imaging and examination of challenging biomedical sample (60 μm thick mouse brain tissue slice). Physical minimization of coherent noise (up to 50%) was positively verified, while preserving optimal spatial resolution of phase and amplitude imaging. Coherent noise removal, ensured by proposed RD-LDHM method, is especially important in biomedical inference, as speckles can falsely imitate valid biological features. Combining this favorable outcome with large field-of-view imaging can promote the use of reported RD-LDHM technique in high-throughput stain-free biomedical screening.

## 1. Introduction

Digital holographic microscopy (DHM) [1-3] enables recording and reconstruction of an optical field that has been modulated via scattering, refraction, absorption or reflection, by a micro-object. Basing on fundamentals of digital holography [4-5] valuable object features can be accessed upon amplitude (absorptive features) and phase (refractive features) demodulation of recorded hologram and numerical propagation to the plane of focus (if hologram was recorded outside of it). Phase component of retrieved complex optical field is recently of great interest in emerging quantitative phase imaging technology [6-8]. Amplitude maps, especially in multimodal systems merged with fluorescence [9], are also very important, e.g., in biomedical imaging [8,10,11]. Lensless DHM (LDHM) setups [12-14] are especially attractive due to their compact setups and ability to omit limitations connected with employment of microscope objective (determined by numerical aperture, depth of focus, field of view, parasitic reflections/aberrations). Large field-of-view (FOV) is available for small object-camera distances and digital in-line holographic configuration [15], developed on the basis of the very first Gabor holographic approach [16]. Off-axis LDHM setups based on Leith and Upatnieks principles [17] are also available [18,19], however they require additional components to decouple object and reference beam and impose restrictions on object field, once operating in total-shear regime. In-line hologram recording determines twin-image errors [20-23], connected with the lack of phase map available in the hologram recording plane (only intensity is captured). Twin-image effect can be minimized upon phase-shifting [21-23], iterative approaches based on angular [24], spectral [25-28], spatial [29] multi-height [30], learning-based [31] data alteration, however the redundancy of data is crucial, and methods need to be significantly modified to allow for it. Hologram recording step is central as pixel size and matrix total size determine achievable resolution of reconstructed complex field. The reconstruction is performed upon numerical propagation, e.g., by angular spectrum method [32-36], to the plane of focus, often determined automatically using so-called autofocusing approaches [37-42], which can be contemporarily also performed within deep learning frameworks [43,44]. The smaller the pixel is the denser fringes this detector can capture (carrying information about finer object details), once they are physically generated with sufficiently high contrast thanks to acceptable degree of spatio-temporal coherence [45]. Detector size is also of pivotal importance as it determines available FOV and information bandwidth. Hence, large size sensors with small pixels are of great interest in LDHM. Uniquely large FOV and high-throughput enabled very capable solutions based on the paradigm of LDHM with applications in biology [46-48], biomedicine [49-54], environmental monitoring [55], nanoparticle tracking [56] and detection [57], point-of-care diagnostics [58] and general technology [59], to name only some.

Noise in recorded hologram directly affects the quality of is further analysis aimed at the complex field reconstruction: noise transfers to amplitude and phase components and can destroy dense fringes limiting available resolution. Noise is more visible where fringe contrast is smaller, which is also detrimental in terms of object signal recovery. The higher the coherence of illumination the stronger the object signal, thus higher contrast of generated fringes, however coherent noise is augmenting. Coherent artifacts generated by laser illumination sources disturb the hologram reconstruction process and lower the signal-to-noise ratio (SNR) of reconstructed complex field, thus are targeted by several minimization approaches, including limiting temporal coherence using LED [60-64] or filtered white light [65] sources. It prevents the formation of dense fringes, however, and can directly truncate the resolution. Moreover, special propagation routines are often necessary to account for broader spectrum [63]. Shorter wavelength allows for higher spatial resolution, however lower coherence length (for a given FWHM) can limit the ability to generate dense fringes. Spatial coherence alteration is also an active topic of research within the LDHM community, with interesting solutions including extended sources [65,66] and variable-size spatial filters [60]. To account for resolution drop, blurred holograms can be preprocessed using deconvolution [65] or regularization [66]. Interestingly, reported solutions of partial spatial coherence use also partially temporally coherent illumination, thus both advantageous (coherent noise minimization) and somewhat troublesome (limited ability to form high-contrast dense fringes) effects are mixing.

In this contribution we performed the LDHM experimental hardware optimization guided by the quantitative evaluation of the numerically reconstructed phase and amplitude components of the optical field (their noise and spatial resolution) generated when imaging phase and amplitude test targets, respectively. For temporal coherence alteration we deployed 4 different light sources: 2 lasers and 2 super luminescent diodes (SLEDs). It is interesting to note that SLEDs are far less popular than LEDs in DHM and constitute a novelty, to the best of our knowledge, in the LDHM. They offer attractive parameters in terms of high brightness and several-nanometer-wide spectrum. To study the hologram sampling effect, we deployed 4 different affordable low-cost cameras.

As a main novelty we propose an original LDHM method with a light source of high temporal coherence and partial spatial coherence. The aim is to limit the hologram noise factor while preserving the ability to transfer small object details encoded in dense Gabor fringes and allow for high contrast in-line interference without the disadvantageous coherence-related blur. We focus on physically driven enhancement of the hologram, not a numerical one [67]. It is achieved, for the first time to the best of our knowledge, by implementing laser light source and rotating diffuser to the LDHM setup. We term the proposed novel method as RD-LDHM (rotating diffuser lensless digital holographic microscopy). In the lens-based DHM the similar type of the light source alteration have been used as pseudo-thermal light source in the Linnik type reflection quantitative phase imaging unit [68-70] and dynamic speckle illumination if grating-based off-axis DHM [71, 72] We study the RD-LDHM performance in transmission mode for lensless holographic examination of amplitude and phase test targets. Additionally, we corroborate its favorable features imaging challenging biological samples (mouse brain tissue slices) and point out the important possibility to remove coherent artifacts which can fraudulently mimic actual biological features.

## 2. LDHM setup description highlighting experimental optimization path

Figure 1 depicts schematic layout of the LDHM setup. It is important to note that single mode optical fiber is coupled to each source to achieve at this point equally high spatial coherence. To understand the effects of temporal coherence and hologram sampling conditions the system was designed to enable easy switching of the light sources and cameras with the constant arrangement of a point source, object, and camera. Below we list four deployed light sources (with central wavelength $\lambda$, full-width half maximum FWHM parameter to grasp the linewidth, and coherence length calculation):

- Laser 1: CNI Lasers MDL-III-405-20mW, $\lambda$ = 405 nm, FWHM = 23 pm, coherence length 2,27 mm

- Laser 2: CNI Lasers MGL-FN-561-20mW, $\lambda$ = 561 nm, FWHM = 47 pm, coherence length 2,13 mm

- SLED 1: EXALOS EXS210084-02, $\lambda$ = 405 nm, FWHM = 3 nm, coherence length 17,4 μm

- SLED 2: EXALOS EXS210033-03 $\lambda$ = 650 nm, FWHM = 6 nm, coherence length 22,4 μm

and four tested detectors used:

- Camera 1: acA5472-17uc, color matrix, pixel size 2,4 x 2,4 μm, 5496x3672 pixels, FOV = 116 mm$^2$,

- Camera 2: FLIR BFS-U3-120S4M-CS, mono matrix, pixel size 1,85 x 1,85 μm, 4000x3000 pixels, FOV = 41 mm$^2$,

- Camera 3: ALVIUM Camera 1800 U-2050m mono Bareboard, mono matrix, pixel size 2,4 x 2,4 μm, 5496x3672 pixels, FOV = 116 mm$^2$,

- Camera 4: BASLER a2A5320-23umBAS, mono matrix, pixel size 2,74 x 2,74 μm, 5328x3040 pixels, FOV = 121 mm$^2$.

Laser with wavelength 405 nm was coupled to a single mode fiber Thorlabs–P1-405B-FC-1, while the laser with wavelength 561 nm was coupled to a single mode fiber Thorlabs – P1-460B-FC-1. Two SLED sources were coupled into an optical fiber with 105 μm fiber core diameter and NA = 0,22. Employed light sources provided the following illumination power: SLED source with wavelength 650 nm: 3mW, SLED source with wavelength 405nm: 200 μW, laser with wavelength 405 nm: 350µW, laser with wavelength 561 nm: around 500 µW. Due to varying illumination power we decided to set the exposure time for each camera and each measurement to allow for capturing holograms with optimal histogram (nicely distributed over 0-255 gray level range).

The length L (Fig. 1) of the entire setup was set to 26 cm to emulate quasi plane-wave propagation between object and camera. Length $Z_1$ (Fig. 1) is the distance between the point source of light and the object. These two dimensions L and $Z_1$ determine optical magnification in the LDHM setup. To get the best configuration for introducing a rotating diffuser, we have carefully studied amplitude and phase LDHM imaging capabilities obtained using representative set of different sources of light, different detectors, and different locations of the object in the setup (resulting in varying optical magnification). We used phase and amplitude test targets to quantitatively study the noise and spatial resolution of phase and amplitude imaging, respectively. Exemplifying holograms are depicted in Fig. 2, whereas amplitude test target (USAF 1951) hologram is presented in Fig. 2(a) and phase test target hologram is shown in Fig. 2(b). Both holograms have orange rectangles indicating areas studied throughout the paper for resolution and noise assessment.

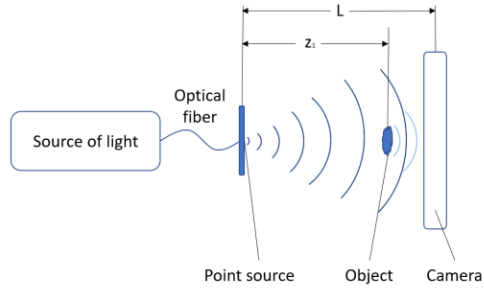

Fig. 1. Lensless digital holographic microscope: schematic layout. L – total length of setup, $Z_1$ – length between point source and object.

It is worth to note that plane of focus has been determined by the DarkTrack [40] metric. Reconstruction process is then conducted upon the object wave $u_{z=0}$, recorded at plane $z = 0$, numerical propagation to a determined focusing distance $z$ using angular spectrum method [32-36]:

$$\tilde{u}_{z=0}(f_x, f_y) = \iint u_{z=0}(x, y) \exp\left[-i2\pi(f_x x + f_y y)\right] dx dy, \quad (1)$$

$$\tilde{u}_z(f_x, f_y) = u_{z=0}(f_x, f_y) \exp\left[-i2\pi z \sqrt{\left(\frac{n}{\lambda}\right)^2 - (f_x + f_y)^2}\right], \quad (2)$$

$$u_z(x, y) = \iint \tilde{u}_z(f_x, f_y) \exp\left[i2\pi(f_x x + f_y y)\right] df_x df_y. \quad (3)$$

In Eqs (1)-(3) tilde denotes Fourier transform, $i$ is an imaginary unit, $(f_x, f_y)$ are spatial frequencies, $\lambda$ stands for the light wavelength, n is refractive index of the medium in which the propagation takes place (in our case refractive index is equal to 1 as we consider free space propagation in air). It is to be noted that in the case of large magnifications and spherical wavefronts appropriate propagation routines should be in place [73,74]. Twin image is an important issue in the in-line holographic microscopy, with capable iterative solutions in place [75]. We decided, however, to maintain our current analysis in a straightforward iterative-free fashion and put emphasis on the hardware optimization and physical noise removal via rotating diffuser.

## 3. Temporal coherence alteration (laser vs. SLED illumination)

Figure 3 shows comparison of amplitude maps reconstructed from holograms of amplitude test target recorded with different light sources (with indicated central wavelengths). The object was placed 2 cm before the Camera 2, thus providing the same object-camera distance for all recordings. Camera 2 was used here, as it has the smallest pixel size. Comparing Fig. 3(a) with Fig. 3(b), depicting reconstruction (amplitude part of reconstructed complex field) of holograms acquired with the the same central wavelength, the achieved resolution is higher for laser than SLED as laser has much higher temporal coherence (coherence length for Laser 1 is equal to 2,27 mm while for SLED 1 it is equal to 17,4 µm). Comparing SLED reconstructions, Fig. 3(b) with Fig. 3(d), higher resolution is obtained for the case of shorter wavelength, as it is theoretically expected. Interestingly, although red SLED 2 has two times wider spectrum (FWHM) it results in coherence length around 22,4 µm which is significantly larger that for violet SLED 1. Nonetheless, much longer central wavelength of SLED 2 plays crucially detrimental role here. In case of laser illumination, Fig. 3(a) and Fig, 3(c), resolving smallest parts of group 7 confirms that the resolution is better than 2,19 µm. To help assess the overall reconstruction contrast we plotted four histograms of obtained amplitude maps, Fig. 3(e).

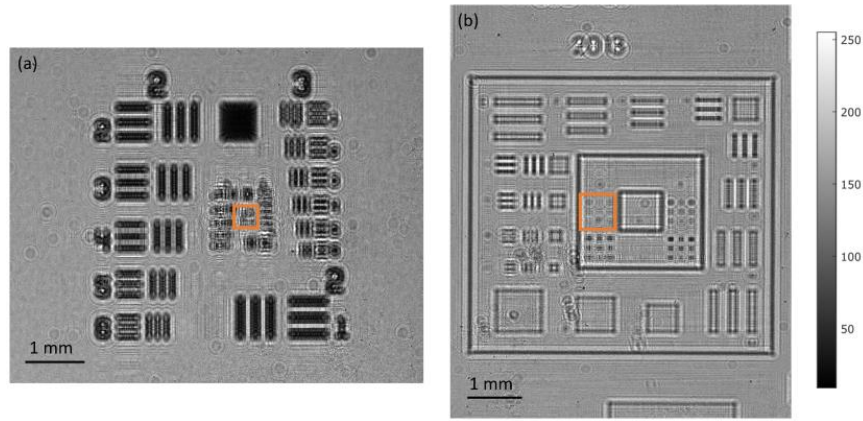

Fig. 2. Exemplifying holograms recorded with standard in-line LDHM setup: (a) hologram of amplitude test target, (b) hologram of phase test target. Orange boxes mark areas of quantitative imaging capabilities assessment, enlarged respectively in following Figures of this manuscript.

Similar results are obtainable for phase test target, see Fig. 4 for phase reconstructions of phase test target holograms and enlarged areas of "R" element. As for the amplitude test target imaging, the best results come with Laser 1, see Fig. 4(a) and corresponding enlarged R-area Fig. 4(e). In conclusion, sources of shorter wavelengths and higher temporal coherence yield better results in amplitude and in phase, due to ability to generate denser fringes of higher contrast. It is to be emphasized that, throughout this contribution, once we study amplitude test target we calculate the amplitude term of the complex optical field and once we examine phase test target we compute the phase term of the complex optical field. Amplitude is depicted in arbitrary intensity units, while phase is depicted in radians. Due to lack of phase in hologram plane and general ambiguity about its global sign we have depicted phase maps after offset correction conducted for visualization purpose. To quantitatively evaluate amplitude and phase noise in LDHM imaging, we calculated the standard deviation (STD) for a given object-free areas marked in Fig. 3(a) and Fig. 4(a) within green squares. STD nicely estimates noise factor in coherent imaging, as has been shown in [76-78].Table 1 presents calculated STD values for amplitude and phase distributions. We can observe noise minimization in amplitude regime related to temporal coherence reduction, however in phase, mainly due to the twin image effect, obtained parameters have similar values for all light sources used. The SLEDs provide less noisy amplitude imaging than lasers, however it is important to bear in mind that resolution is affected as shown in Fig. 4. It is to be emphasized that Table 1 presents also important quantitative experimental evaluation of achievable phase and amplitude resolution (basing on the amplitude/phase test targets imaging) and clearly corroborates the value of optimization of light source employed in LDHM [45].

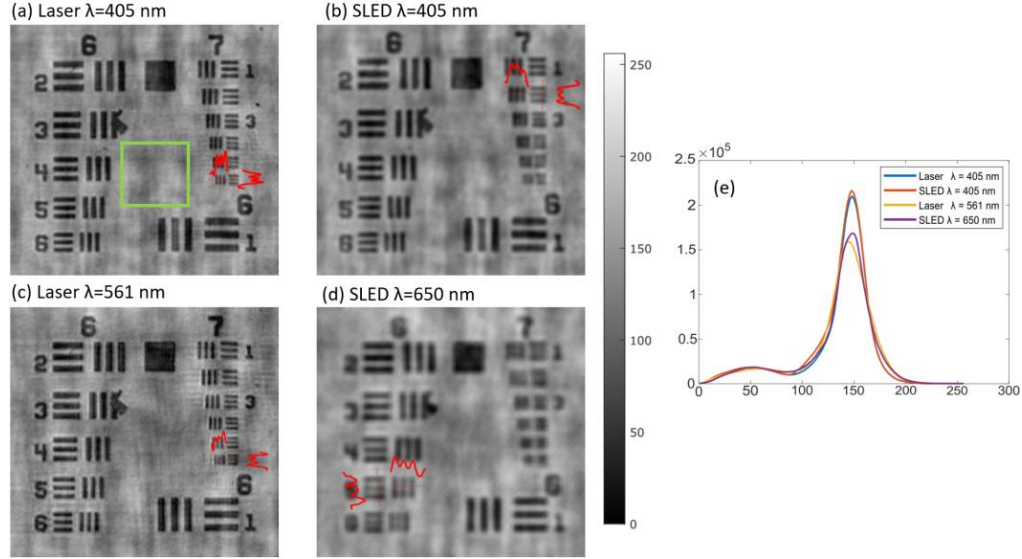

Fig. 3. Amplitude resolution evaluation preformed employing 4 different sources of light. In (a) and in (c) we imaged the finest element (in our test) in the 7th group in which every line has a width equal to 2,19 µm. In picture (b) we have reached the second finest element of the 7th group and the width of the line is 3.48 µm. In (d) the width of the last line we were able to resolve is 4,92 µm. (e) Comparison of the histograms of the images (a)-(d). Green box in (a) marks the area which was used to calculate noise levels (Table 1).

## 4. Hologram physical sampling rate alteration (pixel size and detector size effects)

Sampling of the in-line hologram upon its digital recording plays crucial role in LDHM. The larger the detector area the more scattered information it can capture in terms of the available domain. Within this domain it is crucial to be able to resolve dense Gabor fringes, which encode information about high scattering angles thus allow for transferring fine details of the studied objects. Therefore, the trick is to globally (detector size) and locally (pixel size) capture and sample the in-line Gabor hologram data efficiently. Figure 5 presents the full-FOV holograms recorded employing 4 selected low-cost cameras to pinpoint available global sampling conditions. Figure 6 and Fig. 7 show a comparison of hologram reconstructions for amplitude and phase test targets, respectively. Local sampling conditions are defined by pixel size listed in Section 2 for all four cameras. In Fig. 6 we used a different USAF test target than in Fig. 3, which was a necessity due to unfortunate damage induced to the first test. Comparing amplitude reconstructions presented in Fig. 6(b), Fig. 6(c), and Fig. 6(d) we can observe a tendency to get better resolution using a camera with smaller pixel size. The same relationship can be noticed in images Fig. 7(f), Fig. 7(g), and Fig. 7(h) for phase imaging. Hologram recorded by Camera 1, Fig. 6(a), provided amplitude reconstruction with lowest contrast, which can be easily noticed analyzing histograms presented in Fig. 6(i). For phase imaging, low contrast of Camera 1 hologram resulted in higher phase noise, visible in Fig. 7(a) and Fig. 7(e). Comparing Camera 1 and Camera 3 (with the same pixel size) in amplitude and phase imaging regimes, higher resolution is reached using the camera with a mono matrix. In conclusion for getting the best details in our LDHM, the detector should be monochromatic and pixel size should be minimal.

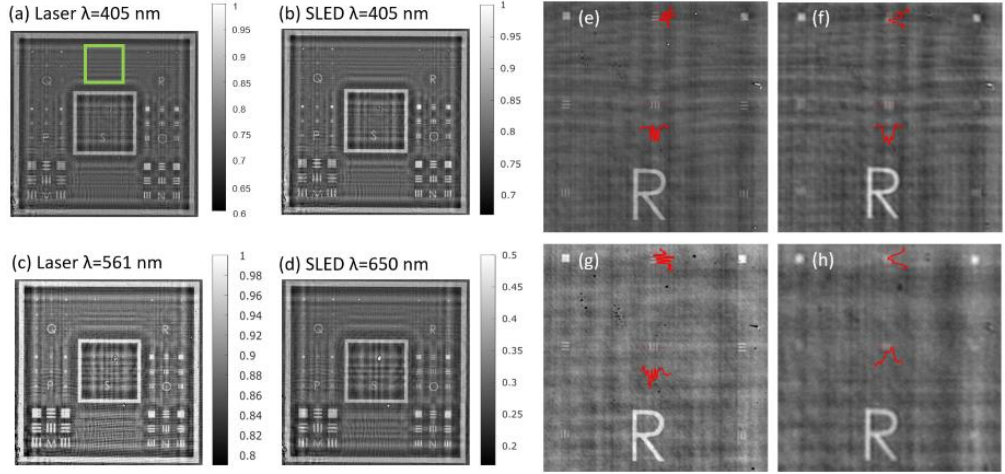

Fig. 4. Images (a)-(d) present phase resolution evaluation preformed employing 4 sources of light. Green box in image (a) marks area which was used to calculate the noise (Table 1). Images (e)-(h) present enlarged areas with "R" element of (a)-(d), respectively.

Table 1. Quantitative evaluation of noise (calculated from Fig. 3 and Fig. 4 within the areas marked by green boxes) and resolution (assessed visually for known widths of elements imaged in Fig. 3 and Fig. 4 for phase and amplitude test targets).

|  | *Laser 1* $\lambda=405nm$ | *SLED 1* $\lambda=405nm$ | *Laser 2* $\lambda=561nm$ | *SLED 2* $\lambda=650nm$ |
|---|---|---|---|---|
| *Noise in amplitude [a.u.]* | 6,6188 | 4,8766 | 7,1751 | 5,2182 |
| *Resolution estimate in amplitude regime [µm]* | 2,19 | 3,48 | 2,19 | 4,92 |
| *Noise in phase [rad]* | 0,0531 | 0,0711 | 0,0676 | 0,0623 |
| *Resolution estimate in phase regime [µm]* | 2 | 3,88 | 2,96 | 6 |

Table 2 sums up the quantitative evaluation of noise related to detector type, calculated similarly to Table 1. In amplitude, the lowest noise was calculated for the color camera, but this camera also provided recordings of holograms with lowest contrast, thus provided significantly lower resolution than the mono camera with the same pixel size (and all mono cameras used throughout this LDHM optimization study). In phase, one can observe very interesting situation. Color camera has the highest noise due to large phase artifacts, e.g., parasitic twin image fringes, and strong noise component related to low quality of the recorded hologram. Camera 4 exhibits large phase noise mainly due to largest pixel size. Other cameras (all mono) have comparable noise parameters in amplitude and phase regimes. It is to be emphasized that Table 2 presented also important quantitative experimental evaluation of achievable phase and amplitude resolution (basing on the amplitude/phase test targets imaging) and clearly corroborates the pivotal importance of optimization of camera employed in LDHM [45].

## 5. Variable optical magnification (object-camera distance)

In the next step, we have compared three reconstructed holograms captured for amplitude and phase test targets, Fig. 8 and Fig. 9, differing in the optical magnification. Variation of optical magnification has been achieved by changing the location of the object in the setup ($Z_1$). Enlarging the object-camera distance comes with increase of optical magnification, clearly seen in Fig. 8 and Fig. 9. We have decided to use Camera 4 as it has the largest detector size, thus provides largest global FOV of recorded holograms. Figure 8 and Fig. 9 contain full FOVs and enlarged parts with fine amplitude and phase test target details to enable resolution assessment. Smallest obtainable optical magnification was dictated by the camera housing, we studied the same set of 3 magnifications for amplitude and phase test target imaging

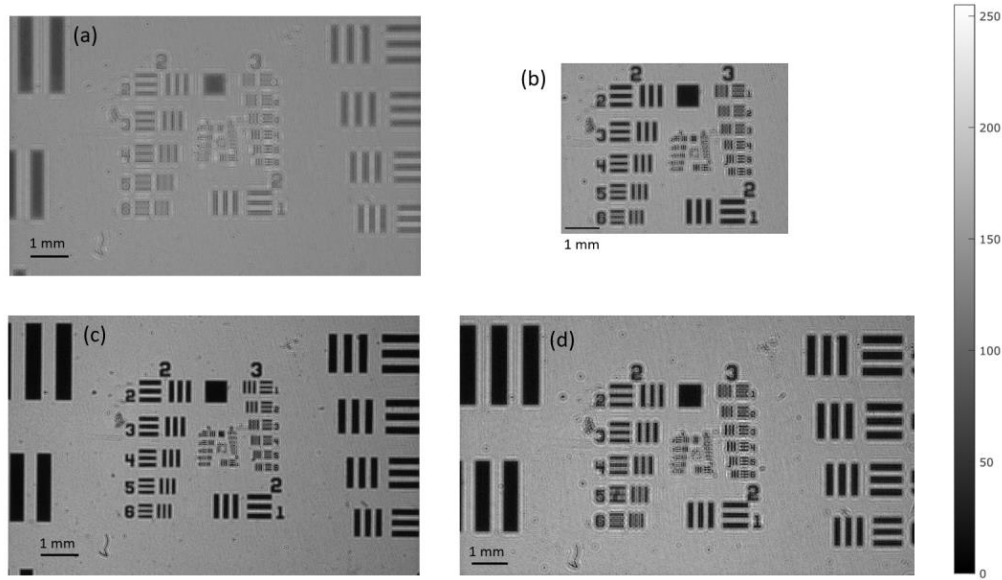

Fig. 5. Comparison of FOV ensured by 4 different low-cost cameras used in our camera-optimization LDHM study: (a) − Camera 1, (b) − Camera 2, (c) − Camera 3, (d) − Camera 4.

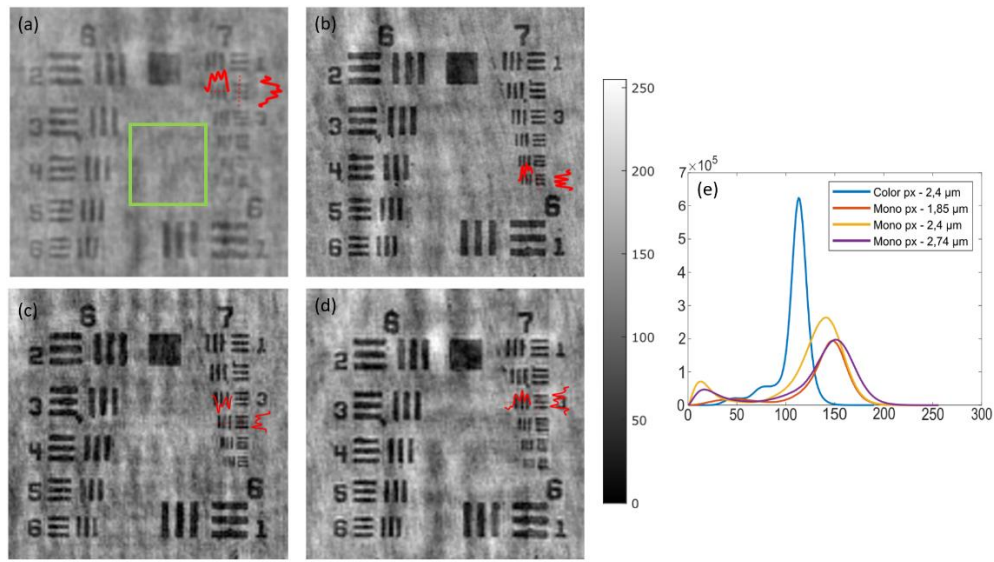

Fig. 6. Amplitude resolution evaluation preformed employing Laser 1 as optimal light source and 4 different low-cost cameras. (a) – Camera 1, (b) – Camera 2, (c) – Camera 3, (d) – Camera 4: enlargements of central part with the smallest amplitude elements of recorded holograms. (e) comparison of the histograms of the images (a)-(d). Green box in image (a) marks the area which was used to calculate the noise (Table 2).

Observing Fig. 8 and Fig. 9 one can clearly note that increase of optical magnification comes with the decrease in resolution, especially visible for horizontal lines as the camera has rectangular FOV with shorter vertical dimension, thus important features of Gabor fringes carrying information about horizontal details are physically truncated and global sampling condition plays a crucial role here. For square camera FOV this effect would not be visible and resolution drop would be isotropic (minus the border effects). Histograms plotted for amplitude indicate that the smaller the distance $Z_1$ the lower the contrast of reconstructed image due to the blur related to diminished global sampling conditions (less object-scattering area is covered by the detector).

Table 3 presented noise parameters calculated similarly to Table 1 and Table 2. Greater magnification results in greater noise, both in amplitude and phase imaging regimes. Resolution is also jeopardized similarly for amplitude and phase imaging. It is to be emphasized that Table 3 presented also important quantitative experimental evaluation of achievable phase and amplitude resolution (basing on the amplitude/phase test targets imaging) and clearly corroborates the pivotal importance of optical magnification employed in LDHM [45], promoting small object-camera distances and largest FOVs. To be able to capitalize on favourable global sampling conditions one should ensure, however, equally good local sampling conditions, thus pixel size is of the essence.

**Table 2. Quantitative evaluation of noise (calculated from Fig. 6 and Fig. 7 within the areas marked by green boxes) and resolution (assessed visually for known widths of elements imaged in Fig. 6 and Fig. 7 for phase and amplitude test targets).**

|  | *Color px – 2,4 µm* | *Mono px – 1,85 µm* | *Mono px – 2,4 µm* | *Mono px – 2,74 µm* |
|---|---|---|---|---|
| *Noise in amplitude [a.u.]* | 5,8097 | 6,6188 | 7,1751 | 6,2847 |
| *Resolution estimate in amplitude regime [µm]* | 3,48 | 2,19 | 2,76 | 3,10 |
| *Noise in phase [rad]* | 0,0736 | 0,0531 | 0,0692 | 0,1128 |
| *Resolution estimate in phase regime [µm]* | 3,88 | 2 | 2,96 | 3,88 |

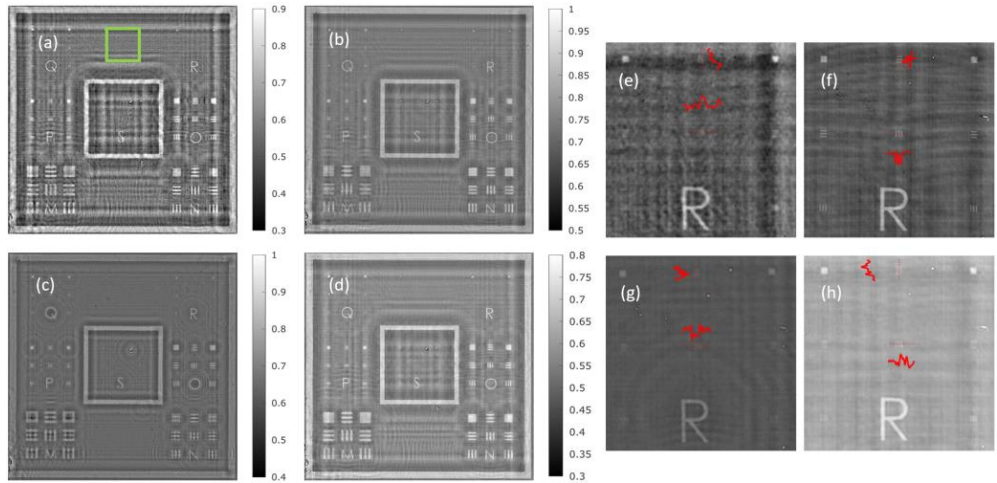

Fig. 7. Phase resolution evaluation preformed employing Laser 1 as optimal light source and 4 different low-cost cameras. (a) – Camera 1; (b) – Camera 2; (c) – Camera 3; (d) – Camera 4. Images (e), (f), (g), and (h) are enlargements of "R" parts of pictures (a)-(d), respectively. Green box in image (a) marks the area which was used to calculate the noise (Table 2).

## 6. Rotating diffuser lensless digital holographic microscopy (RD-LDHM)

Previous sections highlighted the pivotal role of LDHM setup optimization in terms of magnification, camera type and light source type, and showed great important of this step in LDHM system design. Now, we take the LDHM design one step further as our novel RD-LDHM method is poised to physically limit hologram intensity noise and thus decrease amplitude and phase noise by altering spatial coherence of highly temporarily coherent light source. Partial spatial coherence is introduced via rotating diffuser [79-81], which randomizes the phase of focused illumination and cancels out coherent artifacts (high/medium/low frequency speckles, parasitic interferences etc.). Diffusor was made using corundum powder with a grit of 600. Rotation speed was measured using stroboscope light to be around 4000 rpm. The placement of rotating diffusor is very important to keep resolution on the same level, so the diffusor surface with a grit should be very precisely aligned in the very focal point of the light beam created by the microscope objective.

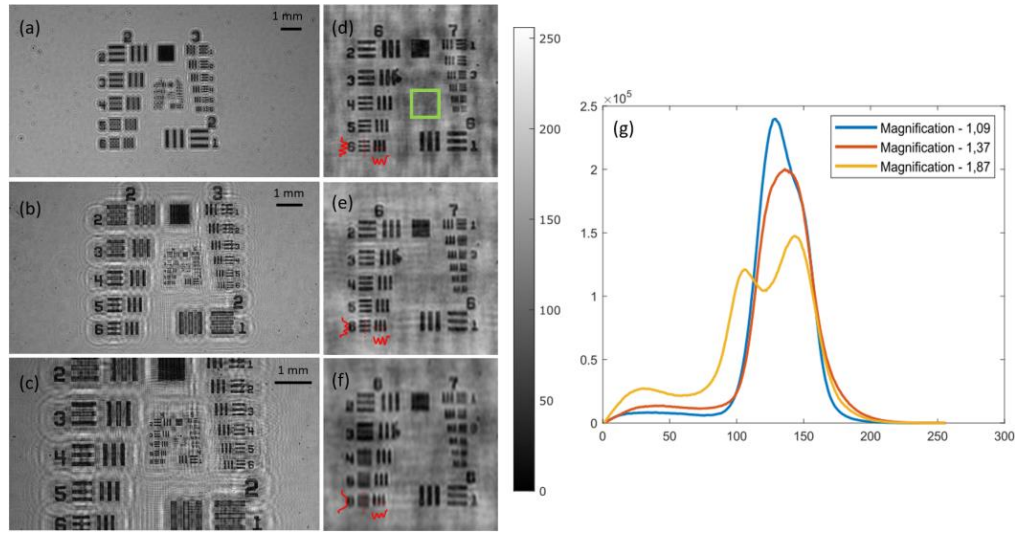

Fig. 8. Amplitude resolution evaluation preformed employing USAF 1951 test target, Laser 1 as optimal light source and Camera 4 as the one with largest FOV. Hologram (a), (b) and (c) were captured with different object-detector distances resulting in varying magnification: (a) 1,09, (b) 1,37, (c) 1,87; Reconstructions (d), (e) and (f) present enlarged central parts (with finest elements) of holograms (d)-(f), respectively. (g) comparison of the histograms of the images (a)-(c). Green box in image (a) marks the area which was used to calculate the amplitude noise levels (Table 3).

Table 3. Quantitative evaluation of noise (calculated from Fig. 8 and Fig. 9 within the areas marked by green boxes) and resolution (assessed visually for known widths of elements imaged in Fig. 8 and Fig. 9 for phase and amplitude test targets).

|  | Magnification – 1,09 | Magnification – 1,37 | Magnification – 1,87 |
|---|---|---|---|
| *Noise in amplitude [a.u.]* | 5,5321 | 6,2873 | 10,5192 |
| *Resolution estimate in amplitude regime [µm]* | 3,48 | 4,38 | 7,81 |
| *Noise in phase [rad]* | 0,0580 | 0,1420 | 0,1666 |
| *Resolution estimate in phase regime [µm]* | 3,88 | 6 | 10 |

Figure 10 shows modifications introduced to the regular LDHM setup to implement, for the first time, pseudo thermal light source working in lensless holographic microscopy in transmission mode. The laser beam is collimated and next focused by the microscope objective on the surface of the rotating diffuser. The focal point of the objective, precisely adjusted to match with the rotating diffuser plate, creates new quasi-point source for RD-LDHM with altered spatial coherence. It is important to note that no collecting module is used [79] which reduces cost and complexity of the setup in the virtue of LDHM vs. classical DHM [82] capabilities.

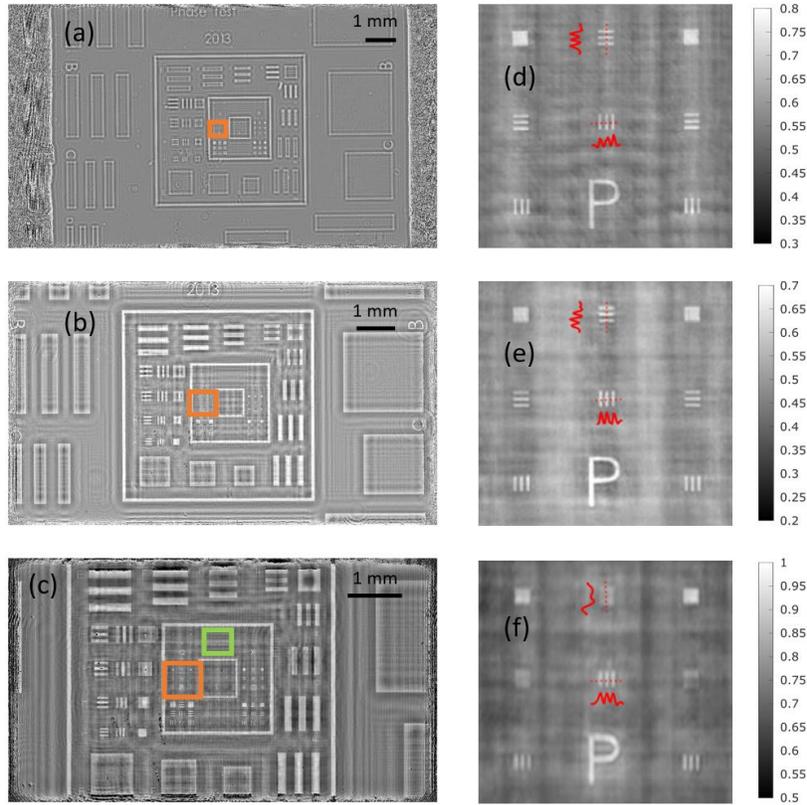

Fig. 9. Phase resolution evaluation preformed employing phase test target, Laser 1 and Camera 4. Holograms (a), (b) and (c) were captured with different object-detector distances resulting in varying magnification: (a) 1,09, (b) 1,37, (c) 1,87; Reconstructions (d), (e) and (f) present enlarged areas around "P" element of phase maps (a)-(c), respectively (within orange boxes).

Comprehensive analysis presented in previous sections allowed us to consciously choose laser light source with 405 nm central wavelength and to employ board-level Camera 3 in order to place the object as close as possible to the detector, in this case around 2 mm. It is important to note that for Camera 3 and object-detector distance equal to 2 mm other light sources executed without rotating diffuser provided very similar results to the once described in previous sections (for 2 cm object-detector distances), thus the grounds for selection of the 405 nm wavelength laser light source stand. Camera 3 has also small pixel and large FOV which additionally promotes its usage corroborated in Sections 3 and 4. Length $Z_1$ was nearly equal to length L so magnification was close to one. In this setup, we have captured holograms of amplitude and phase test targets for two cases: rotating diffuser operating withing the setup, and rotating diffuser fully taken out of the setup.

Figure 11 shows hologram reconstruction results for amplitude object imaged in the RD-LDHM setup with and without rotating diffuser. Figure 11(a) and Fig. 11(b) allow to grasp the effect of rotating diffuser in full FOV of the camera. One can clearly observe cleaner reconstruction and absence of parasitic fringes in the case of operating diffuser. Enlarged areas with finest amplitude features, presented in Fig. 11(c) and Fig. 11(d), help to acknowledge that there is no visible resolution impairment. Direct effect of the rotating diffuser on the captured intensity distribution of in-line holograms can be observed in Fig. 11(e) and Fig. 11(f), where we showed enlarged central areas of both holograms. Figure 11(g) presents plots of histograms of amplitude maps reconstructed via propagation of holograms recorded without and with operating rotating diffuser. One can deduce that conditions of amplitude reconstruction are enhanced in RD-LDHM in comparison with regular LDHM.

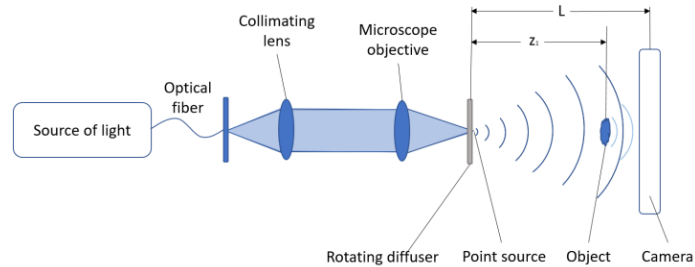

Fig. 10. Modified LDHM scheme with additional rotating diffuser. Setup was built to achieve physical noise minimization. L – total length of setup, $Z_1$ – length between point source and object. Microscope objective – PZO magnification 20x numerical aperture 0,4; collimating lens: Thorlabs f=50mm achromatic dublet (AC254-050-A-ML).

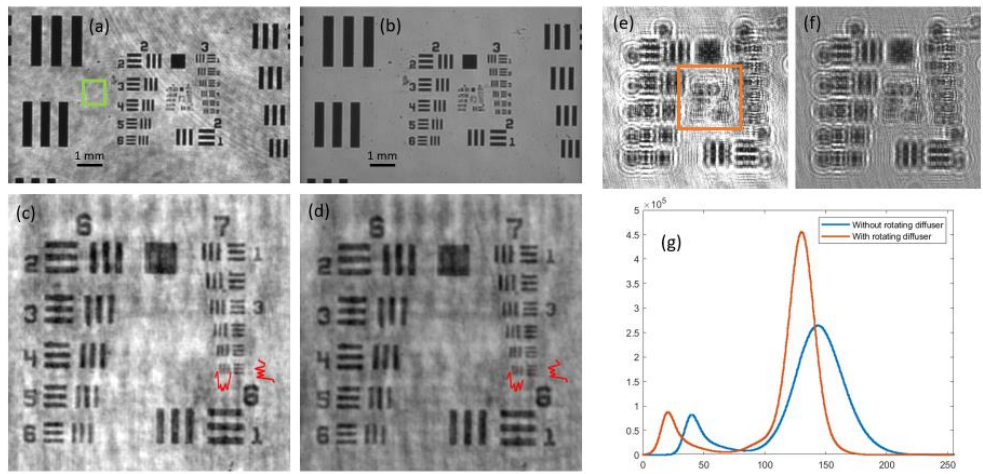

Fig. 11. Evaluation of proposed RD-LDHM in amplitude imaging regime. Full FOV reconstructions: (a) setup without rotating diffuser, (b) set up with a rotating diffuser. Enlargements of central areas with smallest elements: (c) and (d) for diffuser off and on, respectively. Holograms recorded with (e) and without (f) rotating diffuser, orange box marks the central area with smallest details. (g) comparison of the histograms of the images (a)-(b). Green box marks in (a) the area which was used to calculate the noise (Table 4).

Figure 12 presents analogous results for phase test target imaging, both for full FOV and enlarged areas with fine phase features. Smaller phase noise can be observed for RD-LDHM, however the effect is not so visually significant as for amplitude imaging, where scattering has been more impactful and thus alteration of spatial coherence more noticeable. No loss in phase resolution is corroborated, similarly to the amplitude imaging regime. Quantitative evaluation of amplitude and noise parameters, calculated for reconstructions obtained with and without operating diffuser, are stored in Table 4 and help to verify our novel lensless imaging method RD-LDHM. Amplitude and phase noise parameters suggest reduction of noise by 35% for phase and 51% for amplitude reconstructions. It is to be emphasized that Table 4 presented also important quantitative experimental evaluation of achievable phase and amplitude resolution (basing on the amplitude/phase test targets imaging) and clearly corroborates that RD-LDHM achieves optimal imaging conditions in terms of resolution [45], promoting physical denoising via rotating diffuser.

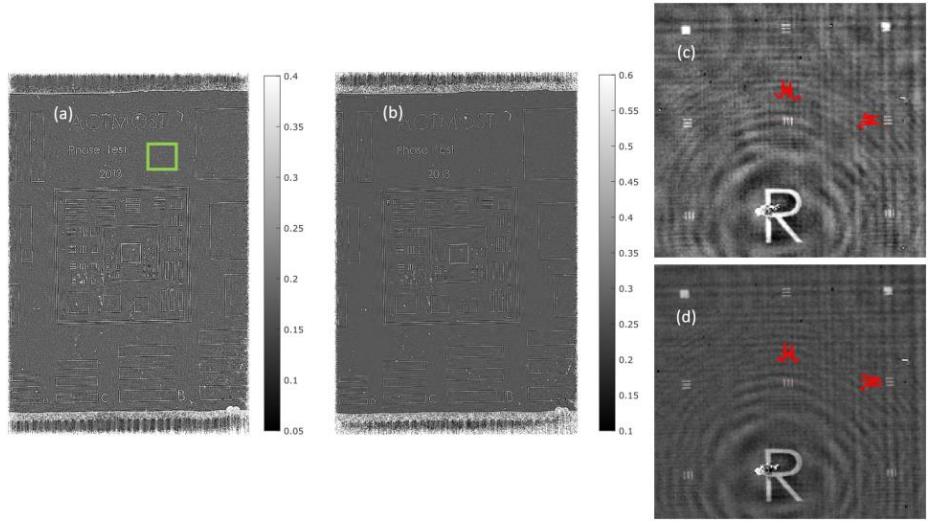

Fig. 12. Evaluation of proposed RD-LDHM in phase imaging regime. Full FOV reconstructions: (a) setup without rotating diffuser, (b) set up with a rotating diffuser. Enlargements of areas around "R" element: (c) and (d) for diffuser off and on, respectively. Green box marks in (a) the area which was used to calculate the noise (Table 4).

## 7. Experimental validation of novel RD-LDHM method for biospecimen imaging

Proposed RD-LDHM method was validated via biomedical sample imaging. Specimen of interest contained brain tissue slice. Brain tissue was collected from an adult C57BL/6J mouse. Animal was first anesthetized with isoflurane and then sacrificed with an injection of a lethal dose of sodium pentobarbital. Next it was perfused using 0.01 M PBS (pH 7.4) and 4% PFA/0.01M PBS. The brain was isolated and postfixed in 4% PFA for 12 hours. Next it was frozen on dry ice and cut on cryostat into 60 μm coronal sections. For imaging, sections were put on poly-D-lysine coated glass slides (Sigma), mounted in Fluoromount-G (Thermofisher) and covered with #1.5 coverslips. For imaging a single coronal section containing the hippocampus was used. No staining was performed. All procedures were performed in accordance with the Animal Protection Act in Poland, Directive 2010/63/EU.

Figure 13 presents amplitude reconstructions of holograms recorded with and without operating diffuser. One can clearly observe severe speckle noise in the case of pure laser light illumination. Once the rotating diffuser is on, those speckles are vanishing upon averaging due to randomized phase in the quasi-point source plane (rotating diffuser plane). Enlarged areas depicted in Fig. 13 serve the purpose of verifying that noise reduction comes with no fine feature ullage. They also help to promote RD-LDHM as stain-free subcellular-specific tissue slice imaging tool. Coherent noise removal, ensured by proposed RD-LDHM method, is especially important in biomedical inference, as speckles can falsely imitate valid biological features (e.g., more obscure areas can indicate special local cell formation, whereas in reality it is pure speckle noise).

In this particular case of brain tissue slice imaging, Fig. 13, fine anatomical details such as parts of the hippocampus formation (blue insert) are visible. It is important to note differences in optical tissue density that show anatomy, i.e., corpus callosum, nerve fiber bundle found beneath the cerebral cortex (orange insert). Combining this favorable outcome with large field-of-view imaging can promote the use of reported RD-LDHM technique in high-throughput biomedical screening. Quantitative noise evaluation shows improvement of 33% (from 11,27 to 7,89 of background standard deviation) which combined with visual judgement forms a very positive corroboration of the RD-LDHM capabilities in large FOV tissue slice imaging.

Table 4. Quantitative evaluation of noise (calculated from Fig. 11 and Fig. 12 within the areas marked by green boxes) and resolution (assessed visually for known widths of elements imaged in Fig. 11 and Fig. 12) for phase and amplitude test targets.

|  | Setup without rotating diffuser | Setup with rotating diffuser |
|---|---|---|
| *Noise in amplitude [a.u.]* | 18,2706 | 8,9429 |
| *Resolution estimate in amplitude regime [µm]* | 2,19 | 2,19 |
| *Noise in phase [rad]* | 0,0639 | 0,0413 |
| *Resolution estimate in phase regime [µm]* | 2,96 | 2,96 |

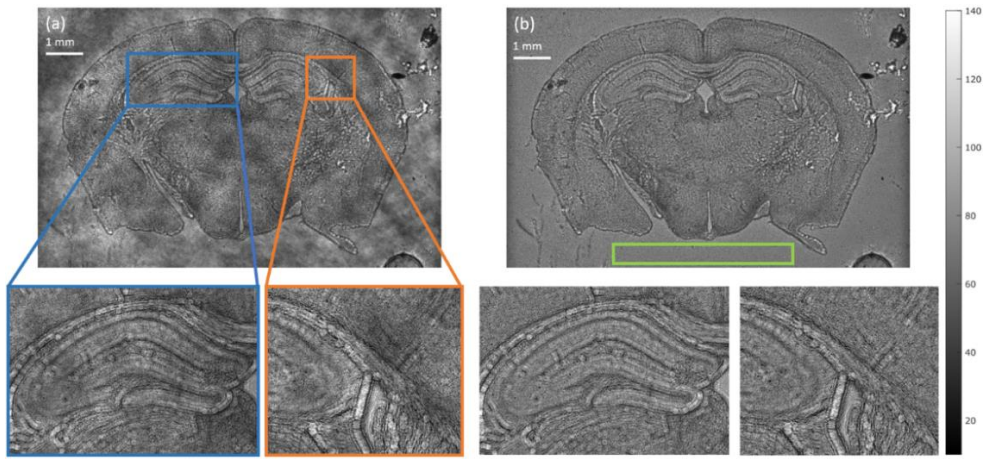

Fig. 13. Mouse coronal brain section examination. (a) full FOV and two enlarged areas obtained without the rotating diffuser, (b) the same presentation of results generated employing novel RD-LDHM approach. Fine anatomical details such as parts of the hippocampus formation (blue insert) are visible. Please note differences in optical tissue density that show anatomy, i.e., corpus callosum, nerve fiber bundle found beneath the cerebral cortex (orange insert). Noise has been evaluated via calculation of STD within the green rectangle area (b) for both imaging modes. Colorbar on the right is depicted in reconstructed amplitude a.u. (gray levels).

## 8. Discussion

First part of the paper consists of the LDHM experimental hardware optimization. Three parameters of pivotal importance influencing the quality (resolution and noise) of phase and amplitude maps reconstructed from in-line holograms recorded with LDHM were examined: light source type, detector type and system optical magnification. The type of the used source of light determines the achievable resolution and the noise component in a sense that a shorter wavelength and narrower spectrum enables greater resolution, while wider spectrum enables noise reduction with a serious risk of detrimental resolution drop (due to diminished ability to generate high contrast and dense Gabor hologram fringes).

Moreover, we have tested SLEDs which were implemented here for the first time in lensless imaging as promising and "more coherent" high-brightness sources than LEDs regularly exploited in LDHM[13,14, 24,31,43, 55,58, 60-66]. We have corroborated that laser sources outperform SLEDs due to more impactful ability to generate dense Gabor fringes of high contrast. We have also shown that the size of pixel and the type of the sensor matrix have enormous impact on the lateral phase and amplitude resolution. Both global (matrix size) and local (pixel size) sampling conditions should be optimized to capture high scattering angles (large camera FOV, short object-camera distance) and resolve dense Gabor fringes (small pixel size). The last parameter that we have examined was the influence of the optical magnification on the amplitude/phase resolution and LDHM imaging noises. The distance between the object and the camera was set to 2 cm as cameras have geometrical limitations with certain sensor covering elements. We have used one board level camera (Camera 3), however, where sensor is directly approachable within 3 mm distance. We decided to present imaging results for this camera for 2 cm object-camera distance, to be able to compare it conscientiously and directly to other cameras. Nonetheless, it is worth to emphasize that achievable resolution and noise levels were similar for Camera 3 with object-camera distances equaled to 2 cm and 3 mm, thus the generality of our analysis is not jeopardized. Our research has shown that, for a given light source and small-pixel camera, the best results are obtained with a magnification close to one, which is achieved by minimizing the distance between the detector and the object (also favorable in terms of high FOV and optimal global sampling conditions).

## 9. Conclusions

In this paper, we have experimentally and quantitatively investigated the optimization of object illumination and hologram recording conditions in the LDHM, with a goal of maximizing the information throughput (increasing the resolution and decreasing the noise). Those very practical aspects of LDHM operation provided valuable insights, versatile in a sense that 4 representative affordable light sources and low-cost cameras were implemented and tested. Important experimentally corroborated conclusions on LDHM imaging conditions optimization facilitated the assembly of a go-to setup and its further novel modification in terms of incorporating rotating diffuser (RD). Proposed RD-LDHM technique, described in detail in the second part of the paper, enabled physical reduction of the hologram intensity speckle noise which directly transfers onto the denoised amplitude and phase imaging, without any noticeable drop in the resolution. It is based on a well-known reduction of the spatial coherence upon modulating the laser beam focal point by a rotating diffuser. It is to be emphasized that this somewhat classical method is innovatively implemented here, for the first time, in LDHM system working in transmission mode offering extremely large FOV, which is the main novelty of the paper. Light source modifications in the LDHM can be considered as a recently popular topic [29, 60-66, 83], and our work aligns within this direction. As a result of introducing the RD-LDHM, beside minimization of the speckle noise, parasitic interference is no longer spoiling the hologram and its reconstruction process. Similarly to the regular LDHM setup, proposed method can be implemented in a compact easy-movable device adding a low-cost microscope objective and a small rotating ground glass diffuser. Thus, we preserved the straightforwardness and compactness of the LDHM imaging device.

We have verified proposed RD-LDHM approach via amplitude and phase test targets examination and highly scattering real-life biomedical sample imaging (60 μm thick mouse brain tissue slice). We showed that new RD-LDHM method enables reduction of amplitude noise by 51%, while phase noise is decreased by 35% for technical test target investigation. Speckle noise reduction in biological sample examination achieved 33% level. The higher the sample scattering and the shorter the wavelength of illumination (which also augments the overall scattering level) the more visible the physical speckle denoising provided by RD-LDHM. Presented experimental results pinpoint the RD-LDHM method as poised to achieve significant noise reduction and ensure preservation of optimal resolution. It is thus envisioned to impact lensless holographic imaging field and allow new biomedical and technical applications where noise elimination is crucial due to generally high overall scattering generating a large share of the coherent artifacts.

**Funding.** This work has been funded by the National Science Center Poland (SONATA 2020/39/D/ST7/03236). Additional partial funding: MS was supported by a National Science Centre grant

(2019/35/B/NZ4/04077), MP was supported by the Foundation for Polish Science under the FIRST TEAM project 'Spatiotemporal photon correlation measurements for quantum metrology and super-resolution microscopy' co-financed by the European Union under the European Regional Development Fund (POIR.04.04.00-00-3004/17-00).

**Disclosures.** The authors declare no conflicts of interest.

**Data availability.** Data underlying the results presented in this paper are not publicly available at this time but may be obtained from the authors upon reasonable request.